\documentclass[%
 reprint,
 amsmath,amssymb,
 aip,
]{revtex4-2}

\usepackage{natbib}
\usepackage{bm}
\usepackage{graphicx}
\usepackage{xcolor}
\usepackage[left]{lineno}
\usepackage{comment}
\usepackage{subfig}
\usepackage{ulem}
\usepackage{booktabs}
\usepackage{multirow}

\newcommand{\degree}{${}^\circ\mathrm{}$}
\newcommand{\degreeC}{${}^\circ\mathrm{C}$}
\newcommand{\degreeCs}{${}^\circ\mathrm{C}$ }

\newcommand*\superR{\textsuperscript{\textregistered}}

\DeclareFontFamily{U}{euc}{}
\DeclareFontShape{U}{euc}{m}{n}{<-6>eurm5<6-8>eurm7<8->eurm10}{}
\DeclareSymbolFont{AMSc}{U}{euc}{m}{n}
\DeclareMathSymbol{\umu}{\mathord}{AMSc}{"16}

\graphicspath{{./Figure/}}

\begin{document}

\preprint{APS/123-QED}

\title{Environmental sub-MeV neutron measurement at the Gran Sasso surface laboratory with a super-fine-grained nuclear emulsion detector}

\author{T. Shiraishi}
 \email{takuya.shiraishi@sci.toho-u.ac.jp.}
\author{S. Akamatsu}
\affiliation{Department of Physics, Toho University, Chiba, Japan}

\author{T. Naka}
\affiliation{Department of Physics, Toho University, Chiba, Japan}
\affiliation{Kobayashi-Maskawa Institute, Nagoya University, Aichi, Japan}

\author{T. Asada}
\affiliation{Università degli studi di Napoli "Federico II", Napoli, Italy }
\affiliation{Istituto Nazionale di Fisica Nucleare, Napoli, Italy}

\author{G. De Lellis}
\affiliation{Università degli studi di Napoli "Federico II", Napoli, Italy }
\affiliation{Istituto Nazionale di Fisica Nucleare, Napoli, Italy}

\author{V. Tioukov}
\affiliation{Istituto Nazionale di Fisica Nucleare, Napoli, Italy}

\author{G. Rosa}
\affiliation{Sezione INFN di Roma, Roma, Italy}

\author{R. Kobayashi}
\affiliation{Graduate School of Science, Nagoya University, Aichi, Japan}

\author{N. D'Ambrosio}
\affiliation{Laboratori Nazionali dell'INFN di Gran Sasso, L'Aquila, Italy}

\author{A. Alexandrov}
\affiliation{Università degli studi di Napoli "Federico II", Napoli, Italy }
\affiliation{Istituto Nazionale di Fisica Nucleare, Napoli, Italy}

\author{O. Sato}
\affiliation{Institute of Materials and Systems for Sustainability, Nagoya University, Aichi, Japan}

\date{\today}

\begin{abstract}
The measurement of environmental neutrons is particularly important in the search for new physics, such as dark matter particles, because neutrons constitute an often-irreducible background source. The measurement of the neutron energy spectra in the sub-MeV scale is technically difficult because it requires a very good energy resolution and a very high $\gamma$-ray rejection power. In this study, we used a super-fine-grained nuclear emulsion, called Nano Imaging Tracker (NIT), as a neutron detector. The main target of neutrons is the hydrogen (proton) content of emulsion films. Through a topological analysis, proton recoils induced by neutron scattering can be detected as tracks with sub-micrometric accuracy. This method shows an extremely high $\gamma$-ray rejection power, at the level of $5 \times 10^7 ~ \gamma/\rm{cm}^2$, which is equivalent to 5 years accumulation of environmental $\gamma$-rays, and a very good energy and direction resolution even in the sub-MeV energy region. In order to carry out this measurement with sufficient statistics, we upgraded the automated scanning system to achieve a speed of 250~g/year/machine. 
We calibrated the detector performance of this system with 880~keV monochromatic neutrons: a very good agreement with the expectation was found for all the relevant kinematic variables. 
The application of the developed method to a sample exposed at the INFN Gran Sasso surface laboratory provided the first measurement of sub-MeV environmental neutrons with a flux of $(7.6 \pm 1.7) \times 10^{-3} \rm{cm}^{-2} \rm{s}^{-1}$ in the proton energy range between 0.25 and 1 MeV (corresponds to neutron energy range between 0.25 and 10 MeV), consistent with the prediction. The neutron energy and direction distributions also show a good agreement.
\end{abstract}

\maketitle



\section{Introduction}

Environmental neutrons are normally a background source for experiments searching for dark matter and neutrinoless double $\beta$-decay in underground laboratories. Therefore, the measurement of their properties including the relative abundance is particularly important for these searches.  For a Weakly Interacting Massive Particle (WIMP)~\cite{WIMP,WIMP2} in the $1 - 10^4$~GeV/c$^2$ mass range, the Maxwell-Boltzman distribution of its velocity in the Milky Way galaxy corresponds to nuclear recoil energies in the $1 - 10$~keV range. Sub-MeV neutrons would produce nuclear recoils with similar energies and therefore their investigation is particularly important for the WIMP search. 

Environmental neutrons in the sub-MeV region have not  been directly measured owing to technical difficulties. In 1988, the measurement of environmental neutrons was carried out by A. Rindi {\it et al.} at the INFN Laboratori Nazionali del Gran Sasso (LNGS)~\cite{GS_neutron}. They used an $^3$He proportional counter, particularly suited for the measurement of thermal neutrons. However, this device detects protons produced by the neutron absorption reaction $^3$He($n$, $p$)T, once the neutrons are decelerated by a moderator. Therefore, a large systematic uncertainty is introduced in the energy resolution by the moderator, which prevents the energy reconstruction for sub-MeV neutrons. Moreover, the detector is not sensitive to the direction of the neutrons since the spatial resolution is not adequate.

We have developed a new direct detection method for neutrons with energies down to the sub-MeV domain~\cite{Neutron}, by using a super-fine-grained nuclear emulsion, called  Nano Imaging Tracker (NIT)~\cite{NIT1,NIT2}. Owing to its unprecedented spatial resolution at the nanometric scale, this device provides the three-dimensional reconstruction of proton tracks induced by the neutron scattering, thus being sensitive to the sub-MeV neutron energy region and providing measurements of both the neutron energy and direction. Moreover, it provides a very high $\gamma$-ray rejection power and it is capable of detecting and measuring neutrons even in an environment with a high $\gamma$-ray rate.

This study is meant to demonstrate the capability of measuring the neutron energy and direction in the sub-MeV domain, by detecting those environmental neutrons at the LNGS surface laboratory. 

In the first part of this work we report about the upgrade of the automated scanning system, to make it faster and collect a larger statistical sample. The   performance of the system in the neutron detection  was carefully measured by using monochromatic neutrons in the sub-MeV region. We then report the results of the  environmental neutron measurements at the LNGS surface laboratory: the neutron flux and its directional distributions in the sub-MeV region are provided. Finally, we discuss the potential of this detection technique for future underground environmental neutron measurements and to search for proton recoils induced by light dark matter scattering.

\section{Detection Technique}
\label{sec:technique}

\subsection{Nano Imaging Tracker}
\label{subsec:nit}

NIT is a super-high resolution nuclear emulsion~\cite{NIT1,NIT2} developed for the NEWSdm experiment~\cite{NEWSdm}, designed to search for dark matter through the direct detection of the induced  nuclear recoils, for the first time with a directional sensitive approach. NIT consists of AgBr:I crystals of several tens of nanometers dispersed in a medium made of gelatin and polyvinyl alcohol:  each crystal acts as the sensor of charged particles. In this study, we used the NIT type with (70~$\pm$~10)~nm AgBr:I crystals, dispersed with a density of about 2000~crystals/$\umu$m$^{3}$, producing an overall mass density of (3.2~$\pm$~0.2)~g/cm$^{3}$.

NIT contains various nuclear targets such as Ag, Br, C, N, O, and H. For the neutron detection, hydrogen acts as the leading target given the larger recoil energy transfer. The hydrogen mass fraction is (1.75~$\pm$~0.30)\%.  
The small size of AgBr:I crystals turns into a large energy deposition per unit length (few 10~keV/$\umu$m) required to sensitize the crystal. This makes NIT insensitive to electrons, except at their stopping point, and thus $\gamma$-rays do not provide a signal track. It makes this neutron detection approach $\gamma$-ray background free.

Nuclear emulsion is usually handled in the form of films, obtained by pouring an emulsion sensitive layer of up to several hundreds micrometers on a mechanical support, known as a base, made of plastic or glass. In this study, we used Cyclo Olefin Polymer (COP) as the base material, due to its low radioactivity from $^{238}$U and $^{232}$Th, and to its high light transmittance, particularly important in the observation at an epi-optical microscope. For the COP base, ZEONOR\superR by ZEON Corporation was selected. The maximum size of the COP base is 120~mm~$\times$~100~mm with a thickness of 2~mm. The NIT emulsion was purified with a 0.22~$\umu$m PES filter (Millex\superR -GP from the Merck company) to remove dust, and it was poured as a 65~$\umu$m-thick sensitive layer on a COP base of 100~mm $\times$ 80~mm size. A thin gelatin layer with 40~nm silver nanoparticles dispersed was applied to the top and bottom as a marker to recognize the emulsion layer.

The sensitization and development process of NIT is similar to that already described in a previous neutron study~\cite{Neutron}. However, due to the larger thickness used for this work, during the fixing treatment at room temperature, NIT samples were soaked for approximately 1.5 hours until the dissolution was confirmed by eye inspection. After this treatment, NIT get shrunk by a factor (0.61 $\pm$ 0.04) w.r.t.~the original thickness. This factor is accounted for during the analysis at the microscope.

\subsection{Three Dimensional Sub-Micrometric Tracking System}
\label{subsec:tracking}

\begin{figure*}
  \includegraphics[width=16cm,bb=0 0 800 500]{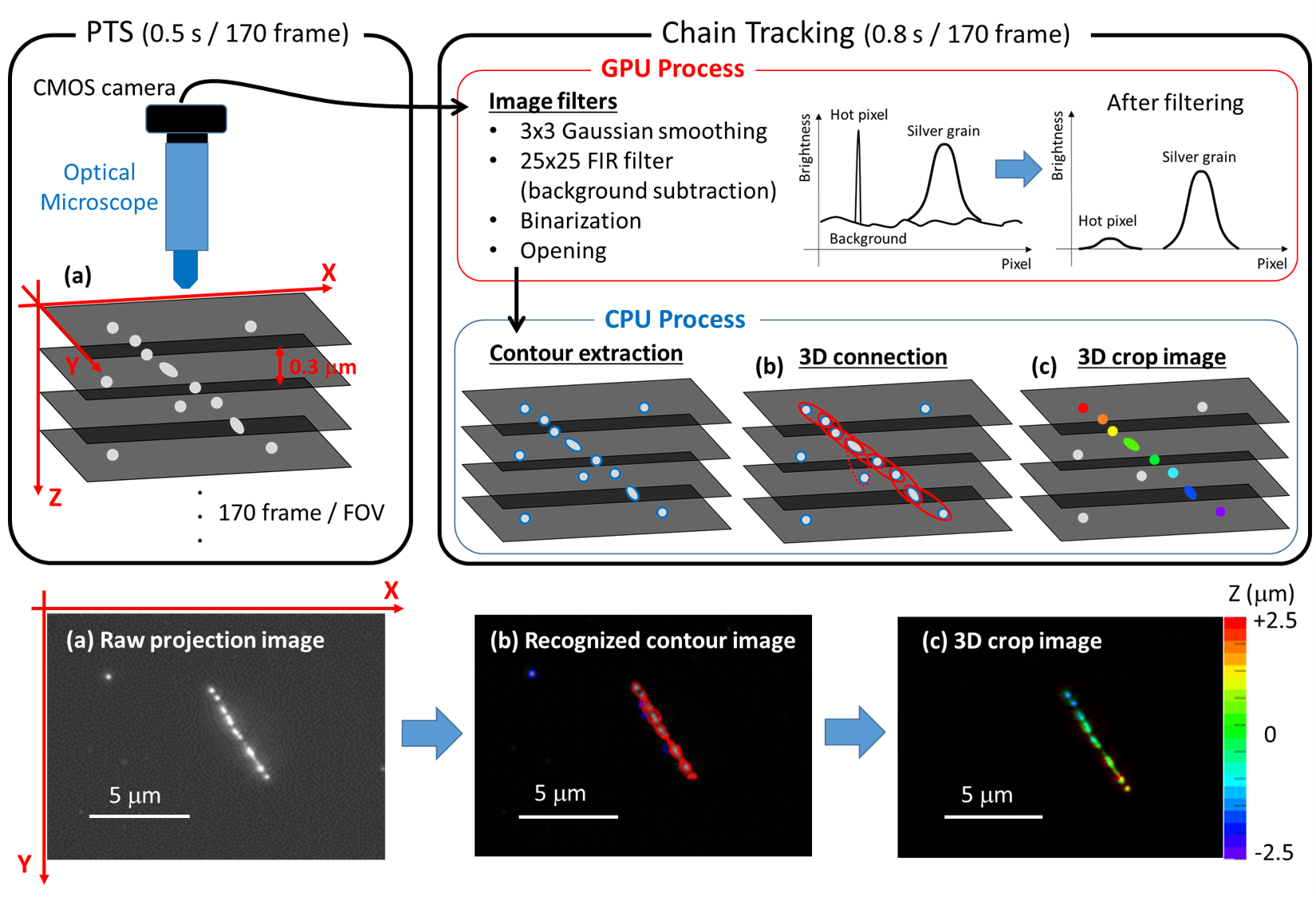}
  \caption{Analysis flow of the 3D track reconstruction algorithm, denoted as Chain Tracking, for tomographic images acquired by the PTS. a) Projection in the Z-direction of 170 frames of tomographic images. After performing a Gaussian smoothing and a background subtraction in each frame with a GPU-based processing, the contours of the developed silver grains are extracted and a 3D connection is performed on the CPU. The recognized track is shown in red (b). Finally, a 3D crop image displays the Z coordinate with a coloured scale (c).}
  \label{fig:image_process}
\end{figure*}

For the NIT analysis, we have developed a three-dimensional sub-micrometric tracking method called Chain Tracking~\cite{Neutron}, by using the scanning system denoted as Post Track Selector (PTS)~\cite{PTS-2,PTS_DFT}, as shown in Fig.~\ref{fig:image_process}. The Chain Tracking is a proprietary 3D track reconstruction algorithm for the tomographic image acquired by the PTS. It first creates pairs of neighboring silver grains produced by the passage of charged particles, then recursively connects, with a chain-like structure, all patterns produced by other silver grains falling within the angular and position allowance. It finally selects the longest chain as a track. This enables automated analysis of tracks longer than 2~$\umu$m with a well-assessed detection efficiency. With this cut on the track length, $\gamma$-rays do not produce detectable tracks, because NIT is sensitive to electrons induced by $\gamma$-rays only at their stopping point. However, the chance coincidence of two $\gamma$-rays has to be considered during a long run when the $\gamma$-ray density increases. We made a dedicated $\gamma$-ray exposure by using an $^{241}$Am source with a density of $5 \times 10^7 ~ \gamma/\rm{cm}^2$, equivalent to the amount of environmental $\gamma$-rays integrated along 5 years. No evidence was found for track candidates induced by $\gamma$-rays which excluded the background from this source.

For all candidate tracks detected by the Chain Tracking, the coordinates (X, Y, Z) of the center of brightness for the two most distant developed silver grains are defined as start and end points, such that the 3D track range and direction are calculated thereafter. 

We have upgraded the objective lens of the microscope. Indeed, in the previous setup, the 100$\times$ objective lens showed a pixel of 0.055~$\umu$m, an over sampling compared to the point spread of about 0.25~$\umu$m due to the diffraction limit. Table~\ref{tab:specification} shows the objective lens and camera used in the current microscope setup: this corresponds to a wider field-of-view (FOV) with a lower sampling pitch and a faster scanning speed. Furthermore, the image analysis in the current setup is performed by a GPU (GeForce RTX 2080 Ti) stream processing to accelerate image filtering, rather than by the CPU parallel processing. Consequently, the analysis speed of the PTS with the Chain Tracking system has achieved 250~g/year/machine, instead of 30~g/year/machine~\cite{Neutron}. In addition, the algorithm was upgraded and optimized to reduce the uncertainty on the 3D range measurement due to mis-connections in the automated analysis.

The depth of field determining the accuracy of this optical system in the direction perpendicular to the film surface (Z-direction) is approximately 0.3~$\umu$m. When acquiring a tomographic image in the Z-direction, the optical system moves at a speed of 0.3~$\umu$m/frame to perform continuous imaging with the camera. During scanning, the emulsion shrunk to approximately 40~$\umu$m, and 170 frames (equivalent to 51~$\umu$m) are acquired in the Z-direction for each FOV.

\begin{table}[htb]
\centering
\caption{Upgraded specification of PTS for the Chain Tracking system.}
\begin{tabular}{|c|c|c|} \hline
  & Previous Work~\cite{Neutron} & Current System \\ \hline
  Objective Lens & N.A. 1.45, 100$\times$ & N.A. 1.42, 66.8$\times$ \\
  Camera Pixel Pitch & 5.5~$\umu$m & 7.0~$\umu$m \\
  Pixel Resolution & 0.055~$\umu$m & 0.105~$\umu$m \\
  Number of Pixels & 2048~$\times$~1088 & 2304~$\times$~1720 \\
  Camera Frame Rate & 300~fps & 500~fps \\
  FOV & 112~$\umu$m $\times$ 60~$\umu$m & 241~$\umu$m $\times$ 180~$\umu$m \\
  Image Processor & CPU & GPU \\ \hline
  Scanning Speed & \multirow{2}{*}{30} & \multirow{2}{*}{250} \\
  (g/year/machine) & & \\ \hline
\end{tabular}
\label{tab:specification}
\end{table}

\section{Detector Calibration by Monochromatic Neutron}
\label{sec:calibration}

In this section, we describe the evaluation of detection performance using monochromatic sub-MeV neutrons generated from a fusion reaction at the National Institute of Advanced Industrial Science and Technology (AIST)~\cite{AIST}.

For the recoil protons detected by the Chain Tracking, the three-dimensional range $R$ [$\umu$m] and the scattering angle ${\theta}_{\rm Scat}$ are measured, and the correlation between the proton range and energy ${E}_{p}$ [MeV] in the NIT is approximated as it follows:
\begin{equation}
  \label{eq:proton_energy}
  {E}_{p} \approx 0.045 + 0.539 \times \sqrt{R} - 0.446 \times \sqrt[3]{R} \quad ({\rm MeV}).
\end{equation}
The neutron energy ${E}_{n}$ in elastic scattering with the proton can be derived from the following equation:
\begin{equation}
  \label{eq:neutron_energy} 
  {E}_{n} = \frac{{(m_n + m_p)}^2}{4 m_n m_p} \frac{{E}_{p}}{{\rm cos}^{2}{\theta}_{\rm Scat}} \simeq \frac{{E}_{p}} {{\rm cos}^{2}{\theta}_{\rm Scat}},
\end{equation}
where we used the approximation $m_n \simeq m_p$, with $m_n$ and $m_p$ the neutron and proton masses, respectively.

In a previous work~\cite{Neutron}, we reported that the energy measurement through  Eq.~\ref{eq:neutron_energy} showed an accuracy of $\Delta E_{n, {\rm FWHM}} / E_n = 0.42$ for 540 keV neutrons.
In this study, we have redone the calibration with monochromatic sub-MeV neutrons to check the effect on the measurement accuracy induced by the upgrade of the optical microscope, and to verify the accuracy obtained through the automated measurement by the Chain Tracking algorithm. In addition, NIT detector was kept at low temperature  to suppress thermal noise and prevent the fading of latent image during long-term measurements. Therefore, we have also prepared a new neutron exposure  to check the sensitivity of NIT films to protons at $-$26~\degreeC.

\begin{figure*}
  \centering
  \includegraphics[width=15cm,bb=0 0 900 400]{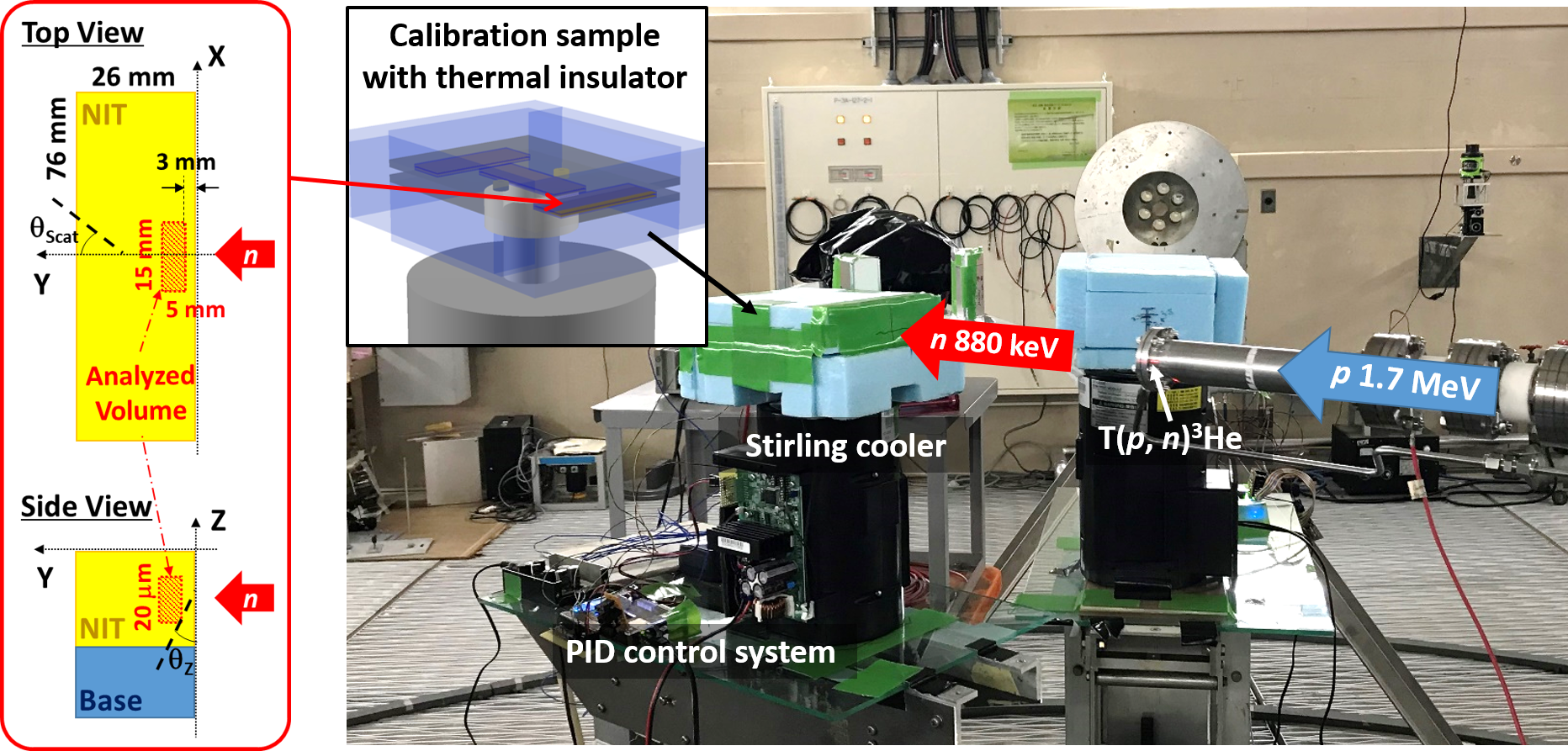}
  \caption{Setup of the exposure of NIT films to a monochromatic neutron beam produced through the T($p$, $n$)$^3$He reaction with a 1.7 MeV proton beam at AIST. }
  \label{fig:AIST2019_setup}
\end{figure*}

We used the monochromatic neutrons produced from the T($p$, $n$)$^3$He reaction by bombarding a tritium-titanium layer evaporated on a 0.5~mm~thick copper backing with a 1.7 MeV proton beam from the 4 MV Pelletron accelerator at AIST~\cite{AIST}. Neutrons emitted in this reaction at an angle of 0\degree~ have an energy $E_n = 880 \pm 20$ keV, and the total flux after 7.88 hours exposure was $(4.75 \pm 0.26) \times 10^7 ~n/{\rm cm}^2$ at the sample location (32 cm away from the neutron source), as measured by the BF$_3$ proportional counter. NIT films were placed in a way to have their surface parallel to the incoming neutrons.  A portable cooling system using a Stirling cooler and a PID control system (see Appendix~\ref{sec:app_Cooling}) was used to keep the temperature stable at $-$26~\degreeCs during the exposure. Fig.~\ref{fig:AIST2019_setup} shows the setup used for the neutron exposure. 
 
\begin{figure}
  \includegraphics[width=9cm,bb=0 0 1200 600]{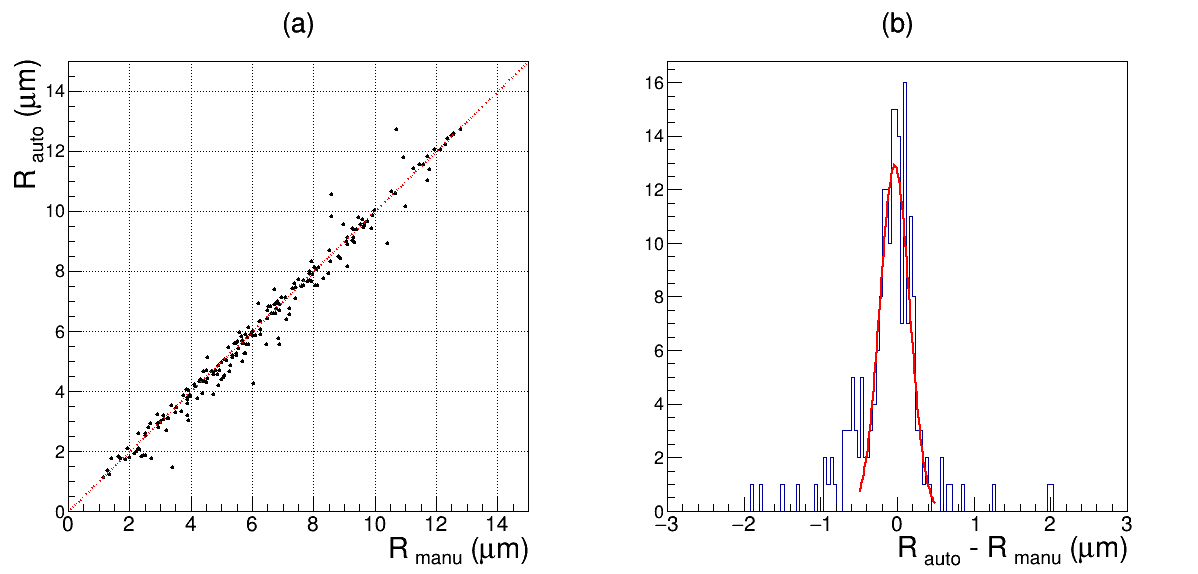}
  \caption{(a) Correlation between manual and automated measurements of the proton range; red dotted line corresponds to equal values. (b) Distribution of the difference between manual and automated range measurements.}
  \label{fig:proton_range}
\end{figure}

In order to evaluate the accuracy in the range measurement by the automatic Chain Tracking algorithm, a comparison track by track with manual measurements was performed, as shown in Fig.~\ref{fig:proton_range}. The automated measurement has an error of approximately 0.2~$\umu$m compared to the manual measurement, which turns into an uncertainty of approximately 20 keV for the proton energy, sufficient to explore the sub-MeV energy spectrum.

In order to evaluate the detection efficiencies, we have made a full simulation of the setup used for the 880 keV monochromatic neutron exposure. The simulation of the neutron propagation relies on Geant4 libraries: G4HadronElasticPhysicsHP and G4HadronPhysicsShielding for the neutron scattering model, and G4EmLivermore for the electromagnetic model. We have included in the simulation the description of all the surrounding materials close to the NIT sample, such as the sample mounting and the Stirling cooler. The neutron flux and its energy spectrum were simulated  for each neutron emission angle, and the tracking pitch for recoil protons in the NIT was set at 0.1~$\umu$m.
In order to avoid the uncertainty associated with the neutron attenuation induced by the scattering, the comparison was done in the proximity of the neutron incident position on the NIT sample. The simulation was normalized to the data, accounting for the actual number of incoming neutrons during the exposure and to the analysed volume.

The number of detected recoil protons in the data was (6330~$\pm$~1280) events, in fair agreement with the predicted value of (5990~$\pm$~70) events. We have estimated for the data a statistical error of 1.3\% and an overall systematic uncertainty of 20.3\% due to the following contributions: 17.1\% to the hydrogen NIT content, 6.5\% to the NIT density, 5.5\% to the neutron fluence, and 6.6\% to the shrinkage factor affecting the actual analysed volume. 

\begin{figure}
  \centering
  \includegraphics[width=7.5cm,bb=0 0 400 1020]{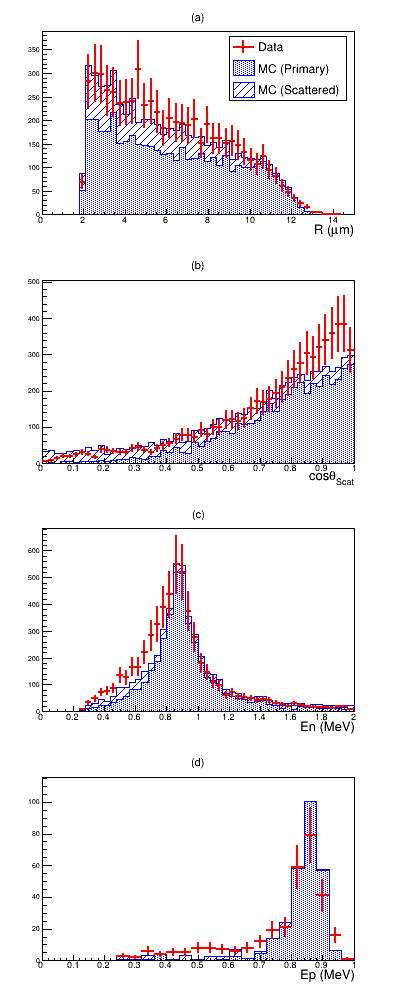}
  \caption{Comparison between data (red) and MC simulation (blue) for the 880 keV monochromatic neutron exposure. For the simulation, primary and scattered neutrons are represented by blue filled and shaded histograms, respectively. (a) Proton range, (b) scattering angle, (c) reconstructed neutron energy using Eq.~\ref{eq:proton_energy} and~\ref{eq:neutron_energy}, (d) recoil-proton energy of head-on collisions.}
  \label{fig:AIST_kinematics}
\end{figure}

Fig.~\ref{fig:AIST_kinematics} shows a data/MC comparison of the measured kinematic variables: proton range ($R$), scattering angle ($\cos {\theta}_{\rm Scat}$), reconstructed neutron energy ($E_n$) and recoil-proton energy ($E_p$) in head-on collisions ($\cos {\theta}_{\rm Scat} > 0.98$). They show a very good agreement both in normalization and in shape. The small excess around $\cos {\theta}_{\rm Scat}$=1 is expected to be due to the scattering from some materials close to the beamline, which was not described in the simulation. Fig.~\ref{fig:AIST_kinematics}(c) reports the neutron energy reconstructed through the recoil-proton energy and the scattering angle, with a peak value at (864~$\pm$~46)~keV, consistent with the exposure energy. The obtained energy resolution is $\Delta E_{n, {\rm FWHM}} / E_n = 0.31$ for 880 keV neutrons, comparable to the value of 0.42 measured in the previous calibration run for 540 keV neutrons~\cite{Neutron}.

\begin{figure}
  \centering
  \includegraphics[width=8cm,bb=0 0 800 600]{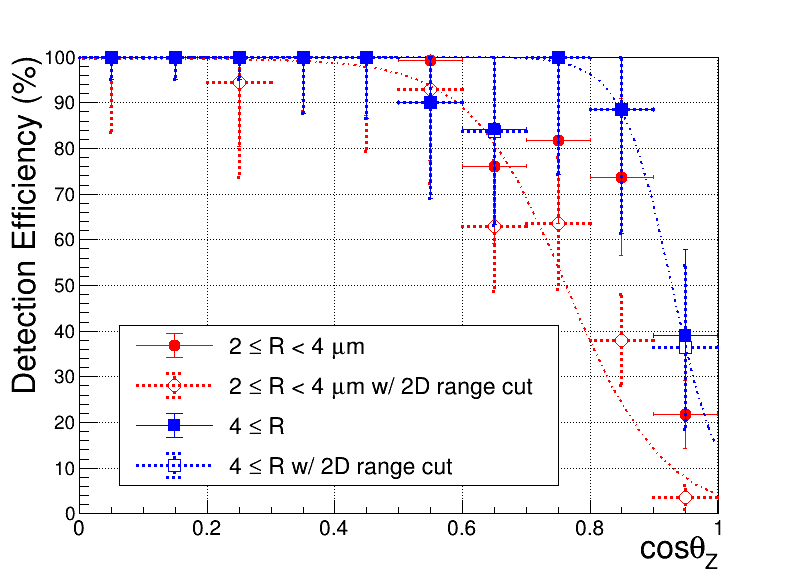}
  \caption{Angular dependency of the detection efficiency for proton tracks induced by 880~keV monochromatic neutron scattering. Open symbols show  the detection efficiency after a 2D range cut is applied, as described in the text.}
  \label{fig:AIST_eff}
\end{figure}

Since most of the protons are scattered at a small angle, the orientation of NIT film adopted in the exposure resulted in a higher detection efficiency. However, as described in Section~\ref{subsec:tracking}, since the accuracy in the Z coordinate is worse than for the other coordinates, a dependency of the detection efficiency is expected on the Z inclination (${\theta}_{\rm Z}$). This is particularly true for short range tracks. The estimated angular dependency of the detection efficiency is reported in Fig.~\ref{fig:AIST_eff}, separately for tracks with ranges within (red) and above (blue) 4~$\umu$m. 

In order to bring the mis-identification of dust events to a negligible level, a displacement between start and end points of the track in the horizontal direction was required to be larger than 1~$\umu$m. This is reported hereafter as a 2D range cut. The detection efficiency gets lower for short vertical tracks once this cut is applied. The angular dependence of the detection efficiency in Fig.~\ref{fig:AIST_eff} was fitted with a Sigmoid function (dash dotted line).

\section{Neutron Measurement at the LNGS Surface laboratory}

We have conducted a run at the LNGS surface laboratory to measure environmental neutrons, given that $\gamma$-rays do not constitute a background in our analysis.

\subsection{Experimental Setup}
\label{subsec:setup}

\begin{figure}
  \centering
  \includegraphics[width=15cm,bb=0 0 1200 300]{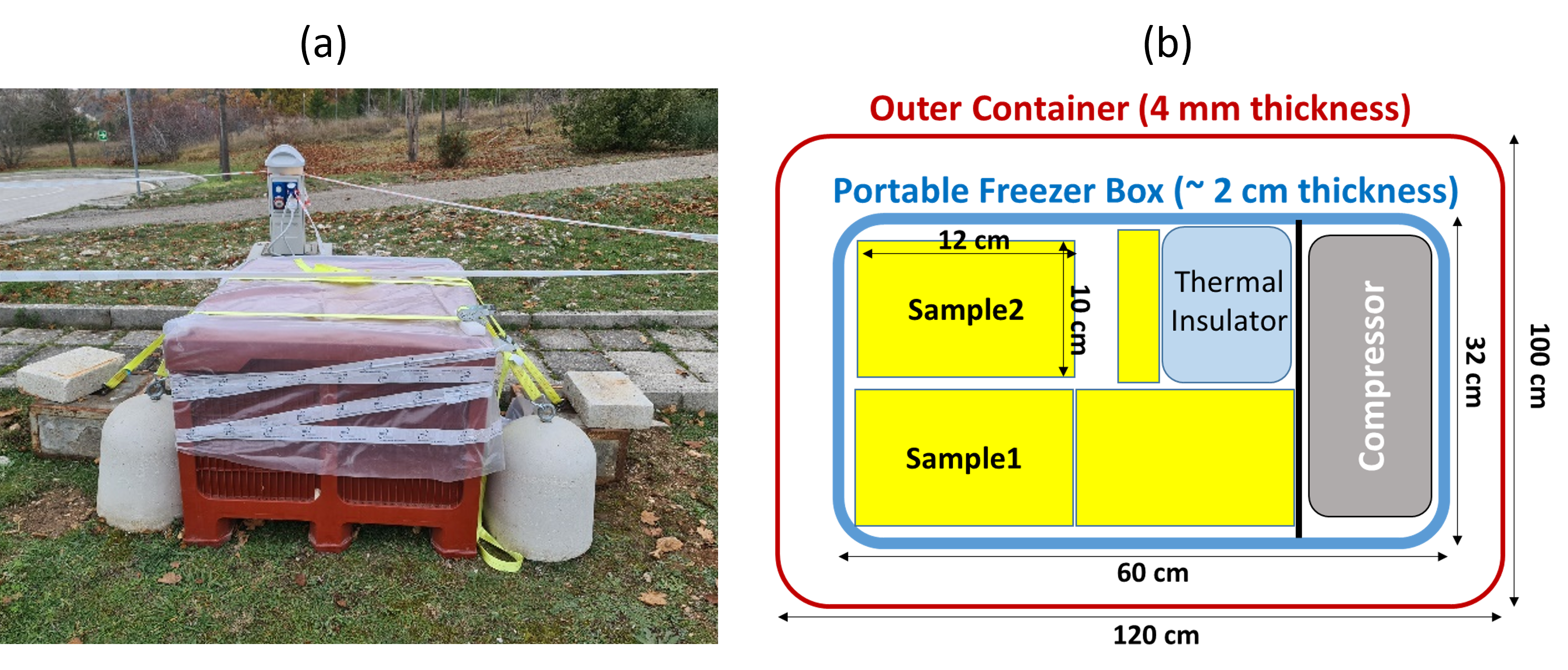}
  \caption{Experimental setup for the environmental neutron measurement at the LNGS surface laboratory. (a) Picture of the setup soon after its installation. (b) Schematic top view of the inner part of the container box.}
  \label{fig:SurfaceRun_setup}
\end{figure}

NIT films were produced at the NEWSdm facility in Hall-F of the Gran Sasso underground. NIT emulsion was produced in the facility, poured on a COP base and dried for 1 day. After that, the films were dipped in a sodium sulfite solution of 0.0397~mol/L for their  Halogen-Acceptor sensitization~\cite{HA} and dried for another day.
These samples prepared underground were transported to the surface laboratory and installed in a portable freezer box located outdoor, as shown in Fig.~\ref{fig:SurfaceRun_setup}. The thickness of the plastic containers was 4~mm for the outer container and approximately 2~cm for the portable freezer box.
Samples were installed for up to 29 days with stable temperature at $-$20\degreeCs to suppress the fading effect as described in Section~\ref{sec:calibration}. 
Table~\ref{tab:setup} shows the details of the two samples used in the measurement. Sample 1 was exposed for two days while Sample 2 was kept for 29 days at the Gran Sasso surface laboratory. The preparation of both samples took 2 days and it was carried out in the underground laboratory. Sample 1 is considered as the reference to study the initial level of radioactivity integrated in the sample.  In order to extract the neutron rate, in the analysis we subtract the rate measured in Sample 1 from the one measured in Sample 2 and consider 27 days as the exposure time.

\begin{table}
\caption{Details of the experimental setup.}
\begin{tabular}{|c||c|c|} \hline
  & Sample 1 & Sample 2 \\ \hline \hline
  Surrounding environment & \multicolumn{2}{c|}{Portable freezer box (outdoor)} \\ \hline
  Altitude & \multicolumn{2}{c|}{1400 m} \\ \hline
  Expected angle-integrated & \multicolumn{2}{c|}{} \\
  flux of atmospheric & \multicolumn{2}{c|}{} \\
  neutron in $0.25 - 10$~MeV & \multicolumn{2}{c|}{$9.0 \times 10^{-3}$ cm$^{-2}$ s$^{-1}$} \\
  (assumed water fraction & \multicolumn{2}{c|}{} \\
  in ground as 20\%)~\cite{EXPACS,EXPACS_ver4.0} & \multicolumn{2}{c|}{} \\ \hline
  Operation temperature & \multicolumn{2}{c|}{$-20$ \degreeC} \\ \hline
  Run start date & \multicolumn{2}{c|}{24 Nov. 2021} \\ \hline
  Preparation time in & \multirow{2}{*}{2} & \multirow{2}{*}{2} \\
  underground (days) & & \\ \hline
  Exposure time (days) & 2 & 29 \\ \hline
  Installation direction & \multicolumn{2}{c|}{Horizontal} \\ \hline
  Analyzed area (cm$^2$) & 46.7 & 99.4 \\ \hline
  Analyzed mass (g) & 0.65 & 1.35 \\ \hline
\end{tabular}
\label{tab:setup}
\end{table}

\subsection{Event Selection}
\label{subsec:selection}

\begin{figure}
  \centering
  \includegraphics[width=13cm,bb=0 0 500 550]{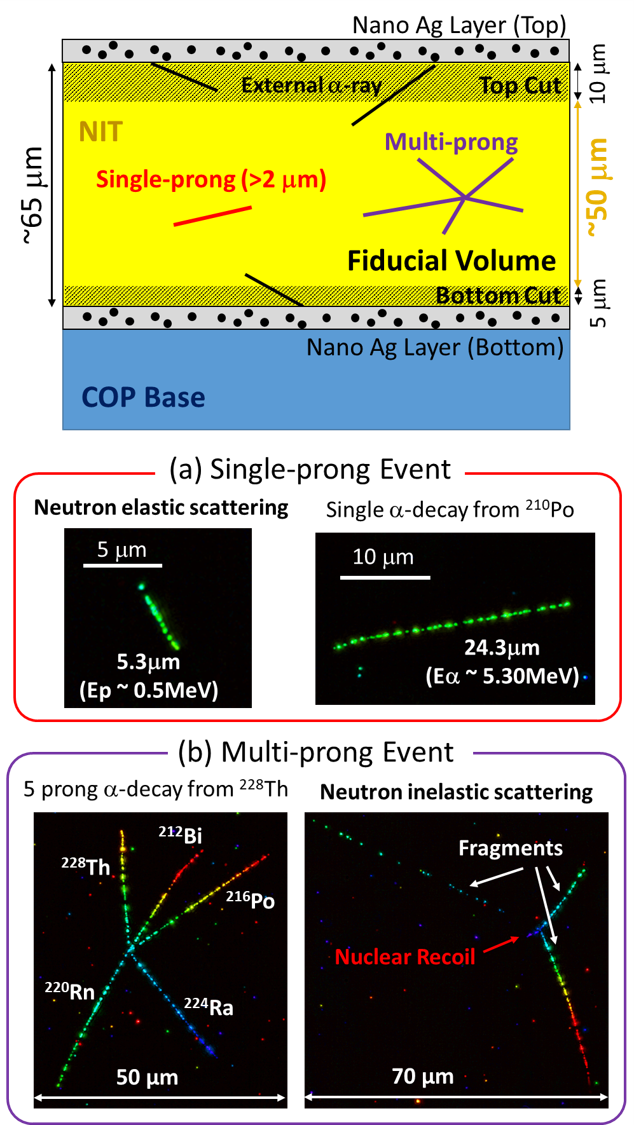}
  \caption{Topological event classification:  (a) Single-prong and (b) Multi-prong events after fiducial volume cut. Coloured scale in images corresponds to the Z coordinate, as shown in Fig.~\ref{fig:image_process}.}
  \label{fig:classification}
\end{figure}

In order to select neutron-induced proton recoil tracks,  we require that both start and end points of the tracks are within the inner fiducial volume which  excludes the 10~$\umu$m from the top and  5~$\umu$m from the bottom of the emulsion. This is meant to  reject external $\alpha$-rays due to $^{222}$Rn from the air and from the $^{238}$U or $^{232}$Th radioactivity in the base materials. Events passing the fiducial volume cut are shown in Fig.~\ref{fig:classification} and are classified as  Single-prong (a) or Multi-prong (b) events, according to the track multiplicity at the vertex.

The intrinsic radioactivity from the $^{238}$U and $^{232}$Th decay chains in the NIT were measured $\gamma$-ray by a germanium detector~\cite{Activity} to be 6~mBq/kg for $^{228}$Th and 0.8~mBq/kg for $^{226}$Ra, and most of the $\alpha$-rays produced show a multi-prong vertex. A typical example is the "Th star~\cite{Th_star}", emitting five $\alpha$-rays in the decay process from $^{228}$Th to $^{208}$Pb. Inelastic scattering events by high-energy neutrons are also observed as Multi-prong, with short-range recoil nuclei and spallation fragments.

\begin{figure}
  \centering
  \includegraphics[width=12cm,bb=0 0 800 450]{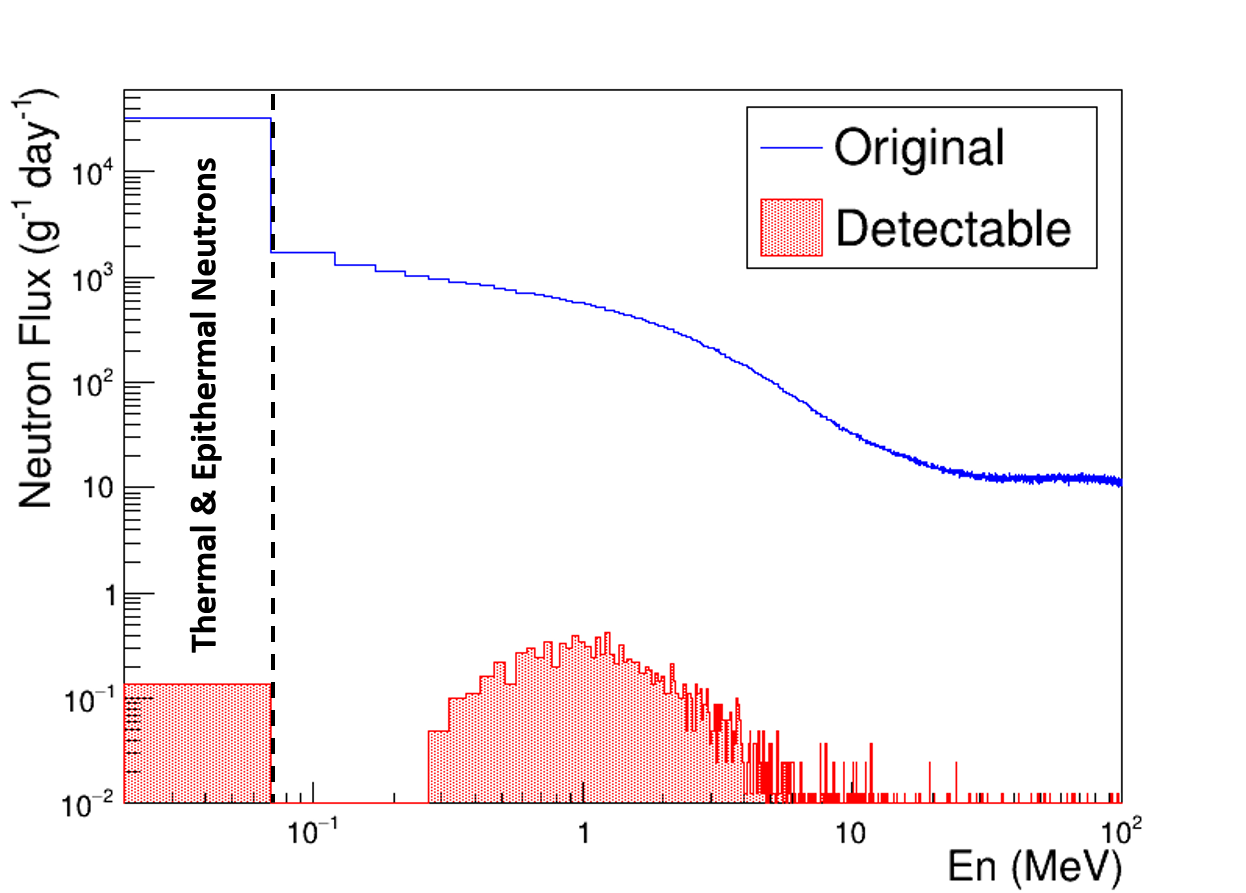}
  \caption{Detectable neutron spectrum in NIT with 1~g~$\cdot$~day exposure at LNGS surface laboratory estimated by a MC simulation based on Geant4. The blue line is the original energy of the incident neutrons, and the red filled histogram is the neutron spectrum accounting for the selection and the detection efficiency in this analysis. Below 100 keV is contribution from the $^{14}$N($n$, $p$)$^{14}$C reaction.}
  \label{fig:neutron_spectrum}
\end{figure}

In this study, we focused on neutron elastic scattering, and only Single-prong events are retained for the analysis. However, $\alpha$-rays might produce a Single-prong event when there is a contamination from $^{214}$Po (7.687~MeV) or $^{210}$Po (5.304~MeV). Their track ranges in NIT are approximately 43~$\umu$m and 24~$\umu$m, respectively (see Appendix~\ref{sec:app_alpha}, \ref{sec:app_MeV}). Therefore, in this study, we set an upper limit for track range of 14 $\umu$m, which corresponds to the proton energy of 1~MeV, and analyze only recoil protons of $2 - 14$~$\umu$m ($0.25 - 1$~MeV in proton energy). The background is therefore negligible in this region. Fig.~\ref{fig:neutron_spectrum} shows the detectable neutron energy spectrum, mostly in the range between 0.25 and 10 MeV, which reflects the cuts applied in the proton range measurement.

In addition, nitrogen contained in the NIT as a mass fraction of (3.7~$\pm$~0.3)\% also produce a small fraction of signal, because the $^{14}$N($n$, $p$)$^{14}$C reaction emits protons with an energy of 0.58~MeV (6.5~$\umu$m in track range) when thermal or epithermal neutrons are captured by nitrogen~\cite{N_np_C}.

\subsection{Result}
\label{subsec:result}
\begin{figure}
  \centering
  \includegraphics[width=8cm,bb=0 0 800 550]{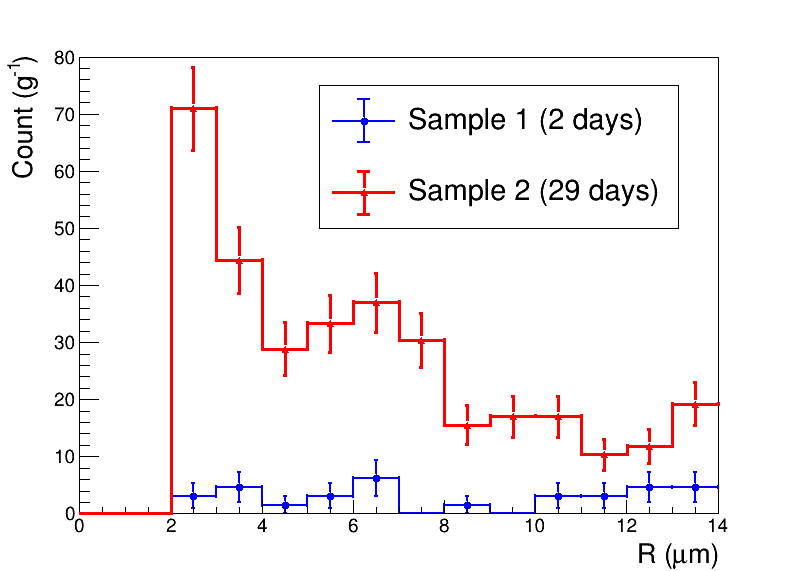}
  \caption{Range distribution of recoil protons in the sub-MeV region for Sample 1 (2 days, blue) and Sample 2 (29 days, red) samples at LNGS.}
  \label{fig:SurfaceRun_range_subMeV}
\end{figure}

Fig.~\ref{fig:SurfaceRun_range_subMeV} shows the range distribution measured in Sample 1 and Sample 2.  The number of detected events was (36~$\pm$~7)~events/g in Sample 1 and (336~$\pm$~16)~events/g in Sample 2, with a significant increase due to the exposure time, as expected. These events are essentially only protons produced in the neutron scattering, given the negligible background.

A MC simulation based on Geant4 was carried out to compare the neutron flux and energy spectrum originated by atmospheric muons.
We considered the neutron spectrum at the LNGS surface laboratory expected by the PARMA~\cite{PARMA} model using the cosmic ray spectrum prediction software of EXPACS~\cite{EXPACS,EXPACS_ver4.0}, which is published by a group of the Japan Atomic Energy Agency (JAEA).
The simulation accounts for a thickness of 4~mm of the container and of 2~cm of the portable freezer box. Neutrons were generated from outside the container considering the Zenith angular dependency predicted by the PARMA model.

\begin{figure}
  \centering
  \includegraphics[width=7cm,bb=0 0 400 850]{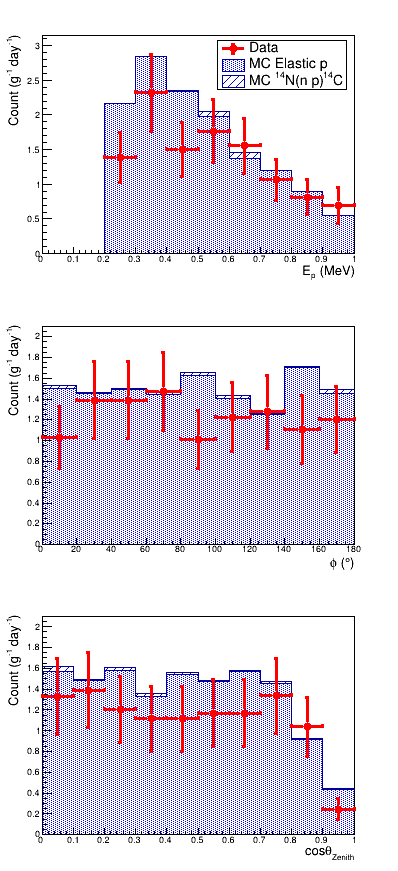}
  \caption{Sub-MeV neutron measurement results after subtracting the data of Sample 1 from Sample 2 for an equivalent exposure of 27 days. For the MC simulation, neutron signals of elastic scattering and $^{14}$N($n$, $p$)$^{14}$C reaction are represented by blue filled and shaded histograms. The detection efficiency reported in Fig.~\ref{fig:AIST_eff} was applied to the MC simulation. (a) Proton energy spectrum, (b) plane angle, and (c) Zenith angle.}
  \label{fig:SurfaceRun_result}
\end{figure}

Fig.~\ref{fig:SurfaceRun_result} shows the measured distributions of the recoil proton energy ($E_p$), plane angle ($\phi$), and Zenith angle (cos$\theta_{\rm Zenith}$) in the data Sample 2 and the comparison with the MC simulation.
The data of Sample 1 were subtracted from Sample 2 to obtain an equivalent exposure of 27 days. The number of events in the proton energy range between 0.25 and 1 MeV was found to be (11.1~$\pm$~0.6(stat.)~$\pm$~2.4(sys.))~event/g/day in the data and (13.2~$\pm$~0.4)~event/g/day in the simulation.  The number of detected event is consistent with the  neutron flux predicted by the PARMA model, and the energy spectrum and directional distribution also show a good agreement. Consequently, we obtained the measured neutron flux as $(7.6 \pm 1.7) \times 10^{-3} \rm{cm}^{-2} \rm{s}^{-1}$ in the proton energy range between 0.25 and 1 MeV (corresponds to neutron energy range between 0.25 and 10 MeV).

\section{Prospects}

We plan to extend this measurement to higher energies. This requires to improve the microscope scanning speed, given the lower flux, and to reduce the background from $\alpha$-rays (see Appendix~\ref{sec:app_MeV}). The increase of the scanning speed will also allow to extend the measurement to the neutrons at LNGS underground where the flux is expected to be three orders of magnitude lower than on the surface. The high accuracy of the emulsion in topological analyses allows extending the analysed sample to events with multiple fragments, thus becoming sensitive also to inelastic neutron scattering, relevant for energies of a few hundred MeV.

Since this neutron detection technique uses the hydrogen content of the emulsion, it also paves the way to the search for low-mass DM. Even though DM masses below 1~GeV/c$^2$ are plausible for the galaxy formation mechanism, they have remained unexplored due to technical difficulties. Recently, Cosmic Ray boosted Dark Matters (CR-DMs), i.e.~DM accelerated by collisions with protons and helium nuclei in the galaxy, was suggested as one of the DM investigation methods~\cite{BDM1}. CR-DMs is a natural consequence of the DM interactions with nucleons, as foreseen by the standard WIMP model. They are predicted to have low mass and a speed higher than the escape velocity of the galaxy: their orientation should preferentially be from the galactic center because of their acceleration mechanism~\cite{BDM2}. By applying the neutron measurement technique with NIT, it is possible to search for low-mass dark matter such as CR-DMs with a directional sensitivity.

\section{Conclusion}

For the environmental neutron measurement, we first upgraded the sub-micrometric 3-dimensional tracking system and achieved an analysis speed of 250~g/year/machine. Then, we calibrated the performance of this system through the analysis of a sample exposed to monochromatic 880~keV neutrons at the temperature of $-$26~\degreeCs and we reported a very good agreement of all the kinematic variables relevant for the neutron elastic scattering. The neutron energy, reconstructed by the recoil proton energy and its scattering angle, was measured to be (864~$\pm$~46)~keV at the peak value, and its accuracy was $\Delta E_{n, {\rm FWHM}} / E_n = 0.31$ with the automated measurement accuracy.

We then performed the environmental neutron measurement at the LNGS surface laboratory. The neutron flux in the proton energy range between 0.25 and 1 MeV was measured to be $(7.6 \pm 1.7) \times 10^{-3} \rm{cm}^{-2} \rm{s}^{-1}$, in good agreement with the prediction by the PARMA model. The uncertainty of this measurement is mainly due to the systematic error associated with the hydrogen content in the films which should be measured with higher accuracy for future and more accurate neutron measurements.

We intend to extend this measurement at the LNGS surface laboratory in the MeV region, by lowering the background contamination from the $\alpha$-ray tracks in the production process and by increasing the statistics. We also plan to perform neutron measurements in the LNGS underground laboratory.

\section*{Acknowledgment}

This work was supported by the Japan Society for the Promotion of Science (JSPS) KAKENHI Grant Numbers JP18H03699, JP19H05806, and JP22J01541. This work was also carried out by the joint usage/research program of the Institute of Materials and Systems for Sustainability (IMaSS), Nagoya University. 

This research was carried out in the frame of the STAR Plus Programme, financially supported by UniNA and Compagnia di San Paolo.

Monochromatic neutron source was supported by Dr. Tetsuro Matsumoto and Dr. Akihiko Masuda of the National Metrology Institute of Japan (NMIJ), the National Institute of Advance Industrial Science and Technology (AIST).

\section*{Appendix}
\appendix

\section{Low Temperature Control System}
\label{sec:app_Cooling}

\begin{figure*}
  \centering
  \includegraphics[width=16cm,bb=0 0 900 260]{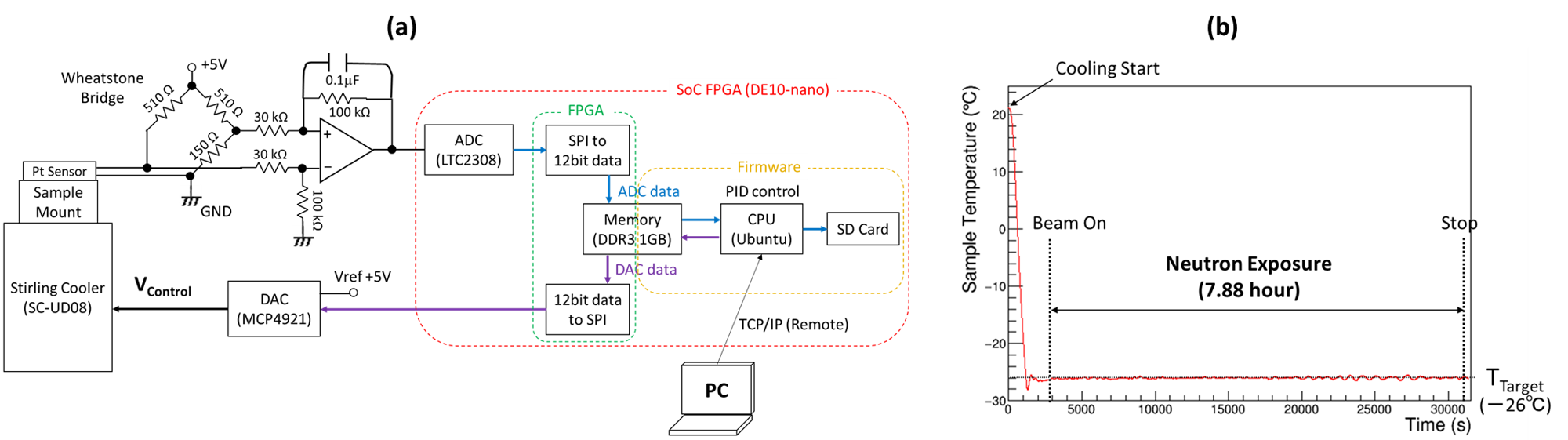}
  \caption{(a) Temperature control system using Stirling cooler (SC-UD08). (b) Temperature profile of a sample during the neutron exposure at AIST. The sample reached the target temperature of $-$26~\degreeCs in approximately 20 minutes. Neutron exposure was carried out for 7.88 hours under stable temperature.}
  \label{fig:cooling_system}
\end{figure*}

We developed a portable cooling system to perform neutron exposure experiments at low temperature. We used SC-UD08 Stirling cooler manufactured by TWINBIRD CORPORATION, which has a cooling capacity of approximately 15 W at $-$100 \degreeCs and 60 W at $-$20 \degreeCs at maximum output, and its output can be controlled by a $1 - 5$ V input voltage. For the control of Stirling cooler, we also developed temperature control system using the SoC-FPGA (DE10-nano), as shown in Fig.~\ref{fig:cooling_system}(a).
First, a platinum resistance thermometer (P0K1.232.6W.B.007), which has good temperature characteristics at low temperatures, is used to convert the sample temperature to a resistance value, then the Wheatstone Bridge circuit converts resistance to voltage. This voltage is digitally converted by the AD converter (LTC2308) when a command is received from the CPU, and the data is written to the shared memory. After the digital conversion, the CPU accesses the data on the shared memory to monitor the current temperature, and determines the control voltage by PID control as described below and sends the data to the DA converter (MCP4921) to set the control voltage of the Stirling cooler.

In the PID control, the voltage for the Stirling cooler ($V_{\rm Control}$) is determined by the following equation with the target temperature $T_{\rm Target}$ and the temperature $T(t)$ at time $t$.
\begin{equation}
  \label{eq:pid_control}
  V_{\rm Control} = K_{p} \Delta T + K_{i} \int_{0}^{t} \Delta T dt  + K_{d} \frac{dT(t)}{dt},
\end{equation}
where $\Delta T = T(t) - T_{\rm Target}$, and the coefficients $K_{p}$, $K_{i}$, and $K_{d}$ were obtained by the ultimate sensitivity method~\cite{PID} as 0.18, $7.8 \times 10^{-4}$, and 10, respectively. A series of feedback control by PID is performed at 5 seconds intervals.
Fig.~\ref{fig:cooling_system}(b) shows the actual temperature profile of the NIT sample during the 880 keV monochromatic neutron exposure at AIST.

\section{Accuracy of $\alpha$-ray energy measurement}
\label{sec:app_alpha}

\begin{figure}
  \centering
  \includegraphics[width=16cm,bb=0 0 800 550]{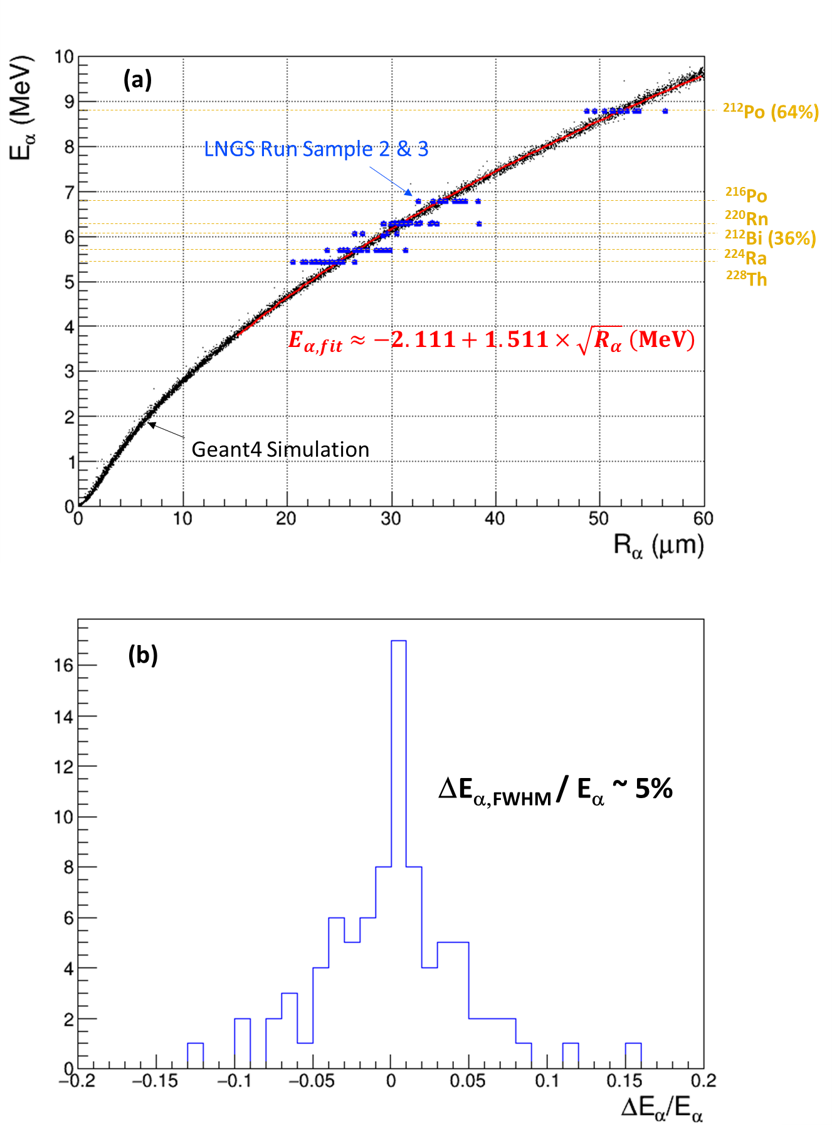}
  \caption{(a) Correlation between $\alpha$-ray range and energy. Blue plots are data using the "Th star" events detected in Sample 2, black plots are the predictions from the Geant4 simulation, and the red line is the best-fit function from the data. (b) $\alpha$-ray energy resolution obtained from the measured range.}
  \label{fig:alpha_calib}
\end{figure}

To identify the $\alpha$-ray source, the energy calibration was performed using the "Th star" events found from Sample 2 of LNGS-run. 5-prong events can be easily identified as "Th star" since they are produced by the decay of $^{228}$Th to $^{208}$Pb  during the run because of their long lifetime. In addition, these tracks can be used to calibrate the correlation between range and decay energy, because it is easy to determine which track  corresponds to which decay. The correlation between measured range and decay energy is shown in Fig.~\ref{fig:alpha_calib} for decays fully contained within the emulsion.  The following  formula can be used to estimate the $\alpha$-ray energy ($E_{\alpha}$) in the range between 4 and 10 MeV.
\begin{equation}
  \label{eq:alpha_energy}
  {E}_{\alpha} \approx -2.111 + 1.511 \times \sqrt{R} \quad ({\rm MeV}).
\end{equation}

\section{MeV Region}
\label{sec:app_MeV}

\begin{figure}
  \centering
  \includegraphics[width=7cm,bb=0 0 800 600]{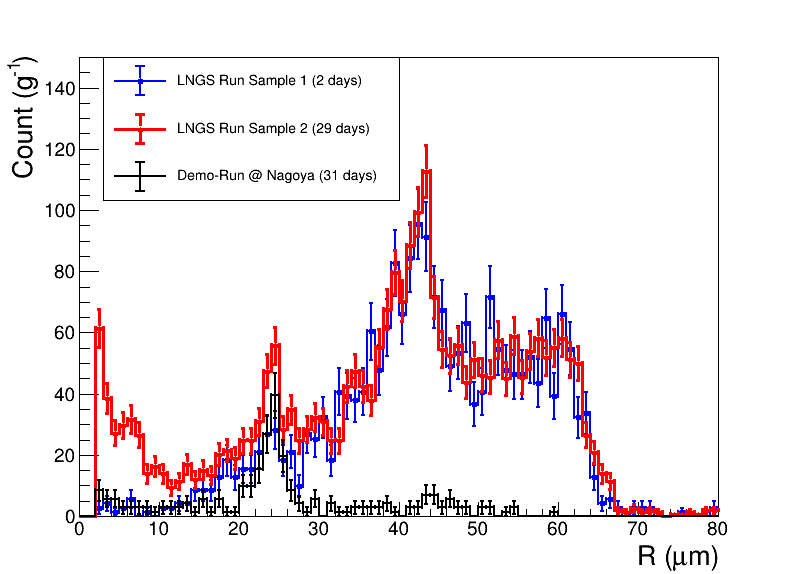}
  \caption{Full range distribution for Sample 1 (2 days, blue) and Sample 2 (29 days, red) samples. A demo-run (black) performed at Nagoya University (Japan) for the background study shows that the background due to longer range events (in the MeV region)  is much lower than at LNGS, due to a different NIT production protocol.}
  \label{fig:SurfaceRun_range_MeV}
\end{figure}

As described in Section~\ref{subsec:selection}, in this study we have focused on the sub-MeV proton energy region to avoid $\alpha$-rays background in the MeV region. The actual range distribution in the MeV region is shown in Fig.~\ref{fig:SurfaceRun_range_MeV}.

As a demonstration run to estimate the amount of background, we had performed similar measurements inside a building at Nagoya University in Japan, where the neutron flux was lower than at LNGS, with 31 days exposure. In the demo-run, a peak at (24.3~$\pm$~0.4)~$\umu$m corresponding to an $\alpha$-ray energy of (5.30~$\pm$~0.08)~MeV was observed and it was identified as $\alpha$-rays from $^{210}$Po with a relatively long half-life. In principle, the MeV region can be analyzed by avoiding this region.

\begin{figure}
  \centering
  \includegraphics[width=7cm,bb=0 0 800 1050]{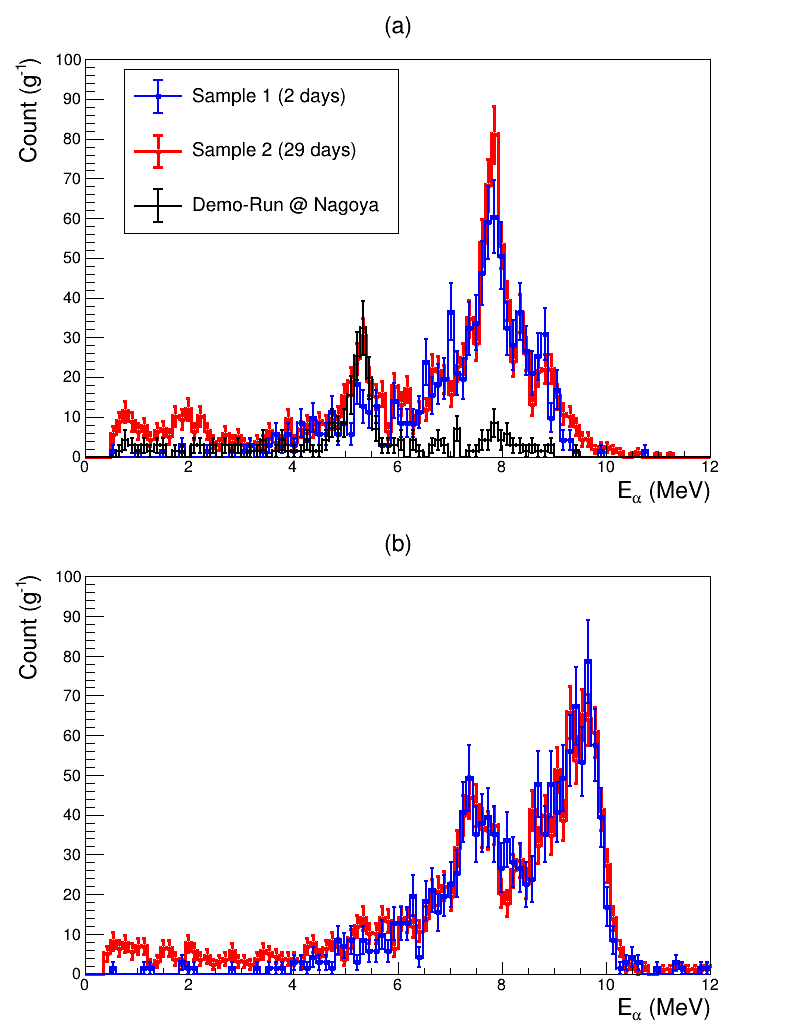}
  \caption{$\alpha$-ray energy spectrum computed by Eq. \ref{eq:alpha_energy} for Sample 1 (2 days, blue) and Sample 2 (29 days, red) samples at LNGS and for a demo-run (black) at Nagoya. Normal (a)  and low (b) brightness tracks.}
  \label{fig:SurfaceRun_alpha}
\end{figure}

However, for the LNGS run reported in Fig.~\ref{fig:SurfaceRun_range_MeV}, we can observe different time-independent contributions in the range between 30 and 65 $\umu$m. We noticed that some of these tracks show low brightness of their reconstructed grains. Fig.~\ref{fig:SurfaceRun_alpha} shows the $\alpha$-ray energy distribution separated in two categories: normal (a) and low (b) brightness.
In the normal brightness track sample, $\alpha$-ray tracks from $^{214}$Po (daughter nucleus of $^{222}$Rn) were observed around (7.81~$\pm$~0.12)~MeV.
In the low brightness track sample, the distribution can be explained by assuming that they originate from $^{214}$Po contamination during the drying process. 
Indeed, during drying, the AgBr:I crystals are more dispersed due to the higher water content which turns into a lower sensitivity to $\alpha$-ray tracks and, hence, to a lower brightness. Their length is also longer because of the lower mass density of the film in this phase. After drying, NIT films shrink in the Z-direction due to the evaporation of water, which turns into a wider range distribution.

We have already established a NIT production method with less contamination of $^{214}$Po at the LNGS underground laboratory which would allow performing neutron measurements with reduced MeV background in the future.


\begin{thebibliography}{9}


\bibitem{WIMP}
G. Steigman and M. S. Turner,
Nucl. Phys. B {\bf 253} (1985) 375.

\bibitem{WIMP2}
D. N. Spergel,
Phys. Rev. D {\bf 37} (1988) 1353.

\bibitem{GS_neutron}
A. Rindi, F. Celani, M. Lindozzi, and S. Miozzi,
Nucl. Inst. Meth. A {\bf 272} (1988) 871-874.

\bibitem{Neutron}
T. Shiraishi, I. Todoroki, T. Naka, A. Umemoto, R. Kobayashi, and O. Sato,
Prog. Theor. Exp. Phys. {\bf 2021} (2021) 043H01.

\bibitem{NIT1}
T. Naka, T. Asada, T. Katsuragawa, K. Hakamata, M. Yoshimoto, K. Kuwabara, M. Nakamura, O. Sato, T. Nakano, Y. Tawara, G. De Lellis, C. Sirignano, and N. D'Ambrossio,
Nucl. Inst. Meth. A {\bf 718} (2013) 519-521.

\bibitem{NIT2}
T. Asada, T. Naka, K. Kuwabara, and M. Yoshimoto,
Prog. Theor. Exp. Phys. {\bf 2017} (2017) 063H01.

\bibitem{NEWSdm}
NEWSdm collaboration,
arXiv:1604.04199v1.

\bibitem{PTS-2}
T. Katsuragawa, A. Umemoto, M. Yoshimoto, T. Naka, and T. Asada,
JINST {\bf 12} (2017) T04002.

\bibitem{PTS_DFT}
A. Umemoto, T. Naka, T. Nakano, R. Kobayashi, T. Shiraishi, and T. Asada,
Prog. Theor. Exp. Phys. {\bf 2020} (2020) 103H02.

\bibitem{AIST}
H. Harano, T. Matsumoto, J. Nishiyama, A. Uritani, and K. Kudo,
AIP Conference Proceedings {\bf 1099} (2009) 915.

\bibitem{HA}
T. Tsutsumi, K. Morimoto, S. Kimura, T. Suzuki, T. Mitsuhashi, K. Kuge, and A. Hasegawa,
J. Imaging Sci. Technol. {\bf 53} (2009) 10507.

\bibitem{EXPACS}
T. Sato,
PLOS ONE {\bf 10} (2015) e0144679.

\bibitem{EXPACS_ver4.0}
T. Sato,
PLOS ONE, {\bf 11} (2016) e0160390.

\bibitem{Activity}
A. Alexandrov, T. Asada, A. Buonaura, L. Consiglio, N. D’Ambrosio, G. De Lellis, A. Di Crescenzo, N. Di Marco, M.L. Di Vacri, S. Furuya, G. Galati, V. Gentile, T. Katsuragawa, M. Laubenstein, A. Lauria, P.F. Loverre, S. Machii, P. Monacelli, M.C. Montesi, T. Naka, F. Pupilli, G. Rosa, O. Sato, P. Strolin, V. Tioukov, A. Umemoto, and M. Yoshimoto,
Astropart. Phys. A {\bf 80} (2016) 16-21.

\bibitem{Th_star}
G. G. Eichholz and F. C. Flack,
J. Chem. Phys., {\bf 19} (1951) 363.

\bibitem{N_np_C}
D. M. Van Patter and W. Whaling,
Rev. Mod. Phys., {\bf 29} (1957) 757.

\bibitem{PARMA}
T. Sato, H. Yasuda, K. Niita, A. Endo, L. Sihver,
Radiat. Res., {\bf 170} (2008) 244.

\bibitem{BDM1}
T. Bringmann and M. Pospelov,
Phys. Rev. Lett., {\bf 122} (2019) 171801.

\bibitem{BDM2}
S. F. Ge, J. Liu, Q. Yuan, and N. Zhou,
Phys. Rev. Lett., {\bf 126} (2021) 091804.

\bibitem{PID}
J. B. Ziegler and N. B. Nichols,
Trans. ASME, {\bf 64} (1942) 759.

\end{thebibliography}
\end{document}